\pdfoutput=1
\documentclass[aps,10pt,nolongbibliography,prl,tightenlines,twocolumn,twoside,showpacs,superscriptaddress]{revtex4-1}
\usepackage[caption=false]{subfig}

\usepackage[hidelinks]{hyperref}
\usepackage{color}
\usepackage{graphicx}
\usepackage{siunitx}
\usepackage{amsmath,amssymb,amsfonts}
\usepackage{bm}
\usepackage{hyperref}
\usepackage{lineno}
\usepackage{changepage}   
\usepackage{slashed}
\usepackage{float}

\usepackage[normalem]{ulem}  

\newcommand\sect[1]{\textit{#1.}---}

\def \be {\begin{equation} }
	\def \ee {\end{equation}}
\def \bes {\begin{subequations} }
	\def \ees {\end{subequations}}

\def \no {\nonumber}
\def \a {\alpha}
\def \b {\beta}

\def \d {\delta}

\def \g {\gamma}
\def \k {\kappa}
\def \o {\omega}

\def \l {\lambda}
\def \m {\mu}
\def \n {\nu}
\def \s {\sigma}
\def \t {\tau}

\def \x {\xi}

\def \pd {\partial}

\def \le {\left}
\def \ri {\right}

\def \<{\langle}
\def \>{\rangle}
\def \+{\dagger}

\def \[{\left[}
\def \]{\right]}

\def \vp {\bm{p}}
\def \vq {\bm{q}}
\def \vx {\bm{x}}
\def \vy {\bm{y}}

\def \ve {\varepsilon}
\def \hn {\hat{n}}

\def \le {\left}
\def \ri {\right}
\def \D {\Delta}
\def \G {\Gamma}
\def \O {\Omega}

\def \<{\langle}
\def \>{\rangle}
\def \+{\dagger}

\def \[{\left[}
\def \]{\right]}

\def \sW {{\cal W}}

\def \sS {{\cal S}}
\def \sT {{\cal T}}
\def \srho 
{{\tilde \rho}}

\def \vz {\bm{z}}

\newcommand{\white}[1]{{\textcolor{white}{#1}}}

\renewcommand{\sout}{\bgroup \color{red} \ULdepth=-.5ex \ULset}

\begin{document}
	\title{Tensor polarization and the dissipative damping of vector meson  in QCD Medium}
	
	\author{Feng Li
	}
	\email{fengli@gmail.com}
	\affiliation{School of Physics and Electronics, Hunan University, Changsha 410082, China} 
	\affiliation{Hunan Provincial Key Laboratory of High-Energy Scale Physics
		and Applications, Hunan University, Changsha 410082, China}
	\affiliation{School of Physical Science and Technology, Lanzhou University, Lanzhou 730000, China}
	\author{Shuai Y.\,F.~Liu
	}
	\email[Corresponding author.\\]{lshphy@hnu.edu.cn}
	\affiliation{School of Physics and Electronics, Hunan University, Changsha 410082, China} 
	\affiliation{Hunan Provincial Key Laboratory of High-Energy Scale Physics
		and Applications, Hunan University, Changsha 410082, China}
	\affiliation{Quark Matter Research Center, Institute of Modern Physics, Chinese Academy of Sciences, Lanzhou 730000, China}

	\date{\today}
	
\begin{abstract}
Unexpectedly large and puzzling spin alignment, and thus tensor polarization, of vector mesons has been observed in heavy-ion collisions. Given that tensor polarization represents a fluctuation of spin, we derive, for the first time, a fluctuation-dissipation relation for tensor polarization, where the polarization is proportional to first-order hydrodynamic gradients (e.g., the shear-stress tensor), with dissipative coefficients depending on the particle's damping properties, as characterized by its spectral function. Employing relativistic hydrodynamics at finite density, we find that dissipative contributions can generate substantial spin alignment. 
We provide illustrative examples (by tuning one coefficients $\alpha_{{\rm sh}}$) that generate a beam energy, $p_T$, and centrality dependence of spin alignment resembling those observed in experiments, offering insights into these puzzling phenomena and demonstrating its potential as a ``spectrometer" for in-medium vector mesons.

\end{abstract}
	
\maketitle

\sect{Introduction}
Spin has played a central role in advancing our understanding of the microscopic quantum world since its discovery. Due to its quantum nature, the measurement of spin is inherently non-deterministic, and we only obtain statistical distributions.  Therefore, statistical quantities such as the mean value of the spin vector $\langle S_i\rangle$(with $i=x,y,z$) and spin fluctuations $\langle S_i S_j\rangle$ are essential for characterizing a particle's spin properties. 
For spin-1/2 particles,  the mean value of the spin vector $\langle S_i \rangle$ is sufficient to determine the complete spin information, i.e., the spin density matrix, of the particle. However, for spin-1 particles, in addition to the vector spin, the symmetric and traceless part of the fluctuation tensor $\langle S_i S_j \rangle$, specifically $\langle S_{(i} S_{j)} - \bm{S}^2 / 3 \rangle$, carries an additional 5 degrees of freedom to characterize the spin density matrix.  This quantity, with proper normalization, is referred to as the tensor polarization of spin-1 particles, and has been investigated in many fields in physics~\cite{Bourrely:1980mr,Haftel:1980zz,Holt:1981qu,Schulze:1984ms,JLABt20:2000uor,Chen:2016moq,Chen:2020pty,ColdAtom}.

In heavy-ion collisions, besides the widely studied vector polarization in experiments~\cite{STAR:2017ckg,STAR:2018gyt,Niida:2018hfw,Adam:2019srw,ALICE:2021pzu} and theory~\cite{Liang:2004ph,Becattini:2013fla,Karpenko:2016jyx,Li:2017slc,Sun:2017xhx,Pang:2016igs,Liu:2019krs,Liu:2021uhn,Fu:2021pok,Becattini:2021suc,Becattini:2021iol,Fu:2022myl,Wu:2022mkr}, the tensor polarization, related to spin fluctuations, of vector mesons has also been proposed to be an observable~\cite{Liang:2004xn}. More concretely, the proposed observable is called spin alignment, which refers to the deviation of  the ``00" component of the spin density matrix from its equilibrium limit: $\delta\rho_{00} = \rho_{00} - 1/3$.  This quantity is directly proportional to the $zz$-component of the tensor polarization: $\delta\rho_{00} \propto \langle 2S_z^2 - S_x^2 - S_y^2 \rangle / 3 = T_{zz}$(quantize along $\hat{z}$), characterizing how fluctuations in the system are unevenly distributed between the transverse $x-y$ plane and the longitudinal $z$-direction.
However, recent measurements in experiments ~\cite{STAR:2022fan,STAR:2008lcm,ALICE:2019aid,Singha:2020qns,ALICE:2022sli} have discovered that the value of the spin alignment is orders of magnitude larger than typical theoretical expectations~\cite{Becattini:2007nd,Becattini:2007sr,Xia:2020tyd,Becattini:2022zvf}, accompanied  by puzzling behaviors. 

Efforts have also been made from various perspectives~\cite{Sheng:2019kmk,Muller2021,Weickgenannt:2022jes,Sheng:2022ffb,Sheng:2022wsy} to understand this puzzle. However, the origin of this unexpected spin alignment remains an open question~\cite{Becattini:2022zvf}. These studies have primarily established non-dissipative relations between spin alignment and other quantities, such as vorticity~\cite{Becattini:2007nd,Becattini:2007sr,Xia:2020tyd,Becattini:2022zvf} or strong electromagnetic fields~\cite{Sheng:2019kmk,Sheng:2022ffb,Sheng:2022wsy}, with the latter being proposed as an explanation in experimental studies~\cite{STAR:2022fan}. However, following Einstein's discovery of the physics of Brownian motion, it is  understood that the fluctuation of quantities can be naturally linked to dissipative processes. Given that spin alignment represents an anisotropic fluctuation of spin, it is natural to ask whether a fluctuation-dissipation relation can be established in this context—a question that has yet to be explored in the study of spin alignment. Therefore, in this work, we present two key findings from our investigation: 

\noindent(1) Theoretically, we establish, for the first time, a fluctuation-dissipation relation for tensor polarization, where novel dissipative contributions are discovered.

\noindent(2) Phenomenologically, the new dissipative effects exhibit promising features that shed light on this puzzling spin alignment observed in experiments.

\sect{Fluctuation-dissipation relation } To further illustrate the fluctuation nature of the tensor polarization and spin alignment, we begin with a simplified example of the spin density matrix $\rho_{ss'}$ (or $\rho$) in the particle's rest frame, where the tensor polarization density $T_{ij}$ can be expressed in terms of the standard $3 \times 3$ spin operators $S^i$ (standard representation~\cite{Li:2025pef}) as $\mathcal T_{i j}\propto-\text{Tr}\{\rho(S_{(i}  S_{j)}-\bm{S}^2/{3} )\}3 n$. Correspondingly, the spin alignment can be expressed as
$\delta\rho_{00}\propto T^{zz}\propto \text{Tr}\{\rho(S_x^2 +S_y^2-2 S_z^2)/3\}$. As shown, both   $T_{\m\n}$ and $\delta\rho_{00}$ can be interpreted as the fluctuation of spin.

To obtain useful formulations for these quantities,  the theory is developed in the following context: vector mesons with momentum $\tilde p^\mu = (\ve_{\vp}, \vp)$ at density $n(\ve_{u})$ propagate in a locally equilibrated hydrodynamic medium, where their tensor polarization $\sT^{\m\n}$ can be characterized by hydrodynamic variables such as inverse temperature $\beta = 1/T$, the flow velocity $u^\mu(x)$, symmetric flow gradients $\xi_{\l\g} \equiv \beta^{-1} \partial_{(\l} (\beta u)_{\g)}$, etc. To study general polarization phenomena in this context, we have developed a comprehensive theoretical framework ``Zubarev response approach", which is detailed in Supplementary Materials (Supp) and in our long theory paper \cite{Li:2025pef}. Especially, in the long paper, we illustrate how our ``Zubarev response approach", rooted rigorously in thermal field theory,  can systematically calculate various polarization phenomena, including the dissipative and non-dissipative contribution on equal footing, and lay out the path to incorporating non-perturbative methods, such as lattice methods and the functional renormalization group.

There are many new dissipative and non-dissipative effects predicted by the ``Zubarev response approach" to the spin alignment problem, where the fluctuation and dissipation relation are naturally encoded. In this letter, we will focus on discussing  one of the main discoveries--shear-induced tensor polarization effects (SITP), using it as an example to highlight how dissipative contributions,  overlooked by the community for decades, can impact the phenomenology of spin alignment physics. Using SITP effect as an example, the uneven spin fluctuation characterized by tensor polarization $\sT^{\mu\nu}$ can be expressed as 
\begin{align}
	\label{eq_Tash}
	\sT^{\mu\nu}
	= \tilde\Delta^{\langle\mu}_\l\tilde\Delta^{\nu\rangle}_\g [\beta n(\ve_{u}) \alpha_{{\rm sh}} \xi^{\g\l}]
\end{align}
which naturally relates to the dissipative processes in-medium through a T-odd dissipative coefficient as
\begin{align}
	\label{eq_ashonly}
	\alpha_{\rm sh}&= \frac{4\ve_{\vp} \pi}{ \beta n(\ve_{\vp})}\int^\infty_0d\o\frac{\partial n(\o)}{\partial \o}  (\o^2-\ve_{\vp}^2) A^2(\o,\vp)\no\\
	& \approx
	-\frac{2\Delta\ve_{\vp}}{\Gamma_{\vp}}+2\frac{\Delta\ve_{\vp}}{\Gamma_{\vp}}\frac{\Delta\ve_{\vp}}{T}+\frac{\Gamma_{\vp}}{2T}\,,
\end{align}
where the spectral function $A(\omega, \vp)$ controls the dissipative damping of a particle's propagation as $D(t,x) \propto \int d\omega\, e^{-i\omega t + \vp \cdot x} A(\omega, \vp)$. The projectors in Eq.~(\ref{eq_Tash}) are defined as $\tilde{\D}^{\m\n}=-\eta^{\m\n}+\tilde{p}^{\m}\tilde{p}^{\n}/m^2\;$ and 	$\tilde{\D}^{\langle\m}_{\;\l}\tilde{\D}^{\nu\rangle}_{\;\g}= \tilde{\D}^{(\m}_{\;\l}\tilde{\D}^{\nu)}_{\;\g}-\tilde{\D}^{\m\n}\tilde{\D}_{\l\g}/3$. 
The second line of Eq.(\ref{eq_ashonly}) employs a quasi-particle approximation, where the $\Delta\ve_{\vp}$ is the in-medium energy shift and is T-even. The $\Gamma_{\vp}$ is the width of the particle, i.e. twice damping rate,  of the particles in-medium, which is the T-odd and is the source of dissipation in this calculation. As shown, the $\alpha_{\rm sh}$ is an odd power of  $\Gamma_{\vp}$ , so it is T-odd and dissipative.

It is worth noting that terms like $\Gamma_{\vp}/(2T)$  do not have mass suppression, allowing them to survive in non-relativistic limits. Consequently, this tensor polarization, especially the SITP effect,  has a potential to be observed in low-energy physics experiments, such as plasma or cold atom systems. Meanwhile, directly extracting the in-medium mass shift and width from the shape of the invariant mass spectra of kaon pairs~\cite{Pal:2002aw} or dileptons~\cite{Li:1994cj} is highly challenging. The observed spectra in experiments~\cite{STAR:2022fan} are dominated by the spin-aligned $\phi$ mesons that survive to kinetic freeze-out with a vacuum spectral function. As shown in Eq.~(\ref{eq_Tash}) and Eq.~(\ref{eq_ashonly}), these in-medium spectral properties are naturally related to the transport coefficients generating the spin alignment, allowing us to  use the spin alignment in reverse as a probe to the spectral properties of the QCD matter. 
In the following, we will use simple setups for the width and mass shift to study how the SITP effect manifests in phenomenology, which reversely illustrates this point.

\sect{Phenomenological setup} We adopt the commonly used freezeout assumption for spin physics~\cite{Becattini:2019ntv,Fu:2021pok,Becattini:2021iol}. 
Specifically, we consider a physical scenario in which the $\phi$ meson, carrying an equilibrated spin alignment, forms as early as the late stages of the QGP~\cite{Shuryak:2004tx,Liu:2017qah}. Depending on conditions such as centrality, beam energy, etc., the fireball evolution varies, leading the spin alignment to effectively freezeout at the QGP phase, mixed phase, or hadronic phase. This contributes to the observed centrality and beam energy dependence (see latter). After this ``spin freeze-out," the spin alignment stays approximately unchanged. The $\phi$ mesons surviving through the kinetic freeze-out will recover the vacuum spectral function but retain a ``memory" of spin alignment from the spin freeze-out. Their decay products no longer undergo re-scattering and will carry all this information to the final observables.
\begin{figure*} [!t]
	\centering
	\includegraphics[width=0.5\columnwidth]{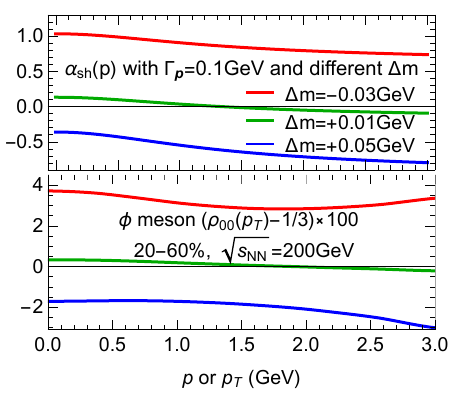}
	\includegraphics[width=0.5\columnwidth]{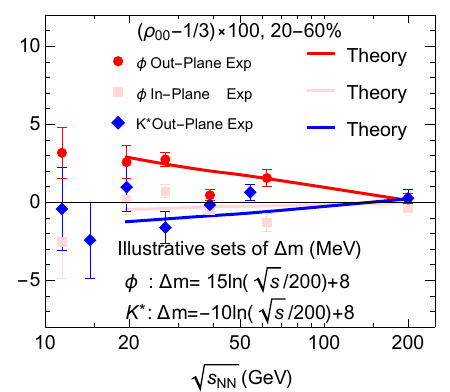}
	\includegraphics[width=0.5\columnwidth]{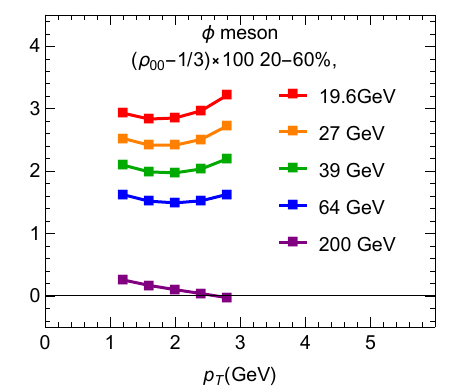}
	\includegraphics[width=0.5\columnwidth]{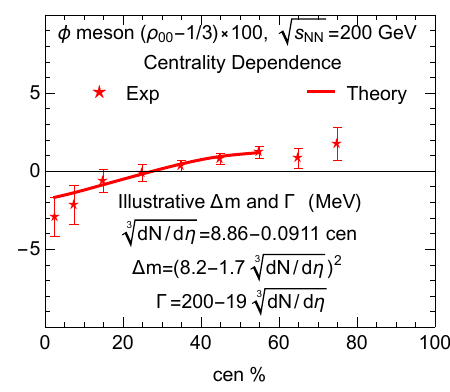}
	\vspace{-0.3cm}
	\caption{
	Compare to experimental data~\cite{STAR:2022fan} by tuning one parameter $\alpha_{\rm sh} (p,T)$($\alpha_{\rm sh} (\D m, \G)$), using the same experimental cuts: (1) $\alpha_{\rm sh}$ for different cases and their $\rho_{00}$; (2) Beam energy dependence; (3) $p_T$ dependence; (4) Centrality dependence.
	}
	\label{fig_ansatz}
	\vspace{-0cm}
\end{figure*}

We employ the standard Cooper-Fry-like formula to characterize the freezeout of spin alignment, where the $\delta \rho_{00}$ can be expressed with tensor polarization $\sT^{\mu\nu}$ as
\begin{align}
	\label{eq_rho-nT}
	\delta \rho_{00}(\hat{n}_{\text{pr}},\vp)=\frac{\int d\Sigma^{\l}p_{\l}\, \sT^{\mu\nu}(x,\vp)\hn_{\mu}(\vp)\hn_{\nu}(\vp)}{d\Sigma^{\l}p_{\l} 3 n(\ve_u)}\,.
\end{align}
The $\Sigma^\lambda$ is the hyper-surface satisfying the freeze-out condition, and $\hat n^\mu$ is a shorthand for $\epsilon^\mu_{s=0}$, which is related to a three-dimensional unit vector $\hat n_{\text{pr}}$ by a Lorentz boost, i.e., $\hat{n}^{\mu}=[\Lambda(\mathbf p).\hat{n}_{\text{pr}}]^{\mu}$, with $\hat{n}_{\text{pr}}$ being the polarization vector $\epsilon^\mu_{s=0,\text{pr}}$ in the particle rest frame. The $\hat{n}_{\text{pr}}$ is taken out/in-plane ($\hat y/\hat x$) direction in this work. 
We utilize the CLVisc~\cite{Pang:2016igs} at finite density~\cite{Wu:2021fjf} for the hydrodynamic simulation at various beam energies and centralities. Unless otherwise stated, the settings are directly inherited from the work~\cite{Wu:2021fjf,Chen:2024xbi} without re-tuning in our event-average qualitative study.
In the following, we will evaluate the spin alignment with $\a_{\rm sh}$ from two setups,  ``physically motivated coefficients" or microscopic ``quark-meson (QM) model coefficients", aiming to extract model-independent features of SITP effects.

\sect{Comparison with data and ``spectrometer" }
\label{sec_unit}
In this section, we tune only one parameter, $\alpha_{\text{sh}}(p, T)$, for the SITP effect and compare the resulting spin alignment with experimental data.  As previously explained, $\alpha_{\text{sh}}(p, T)$ is a function of $\Delta m$ and $\Gamma$. Therefore, we will examine how sensitively the data can, in turn, serve as a ``spectrometer" for vector meson properties.

We first construct a physically motivated  $\a_{\rm sh}$ with the help of Eq.~(\ref{eq_ashonly}). Motivated by Ref~\cite{Rapp:2000pe,vanHees:2007th,Vujanovic:2009wr,Liu:2016ysz,Liu:2017qah} and the QM model discussed later, we introduce some typical momentum dependence using a Lorentz contraction  $\gamma_{\vp}^{-1}=m/\ve_{\vp}\equiv\sqrt{1-v_{\vp}^2}$ factor for the width and its corresponding in-medium mass shifts as $	\G_{\vp}=\Gamma_0/\gamma_{\vp}$ and $\D\ve_{\vp}=\D m/\gamma_{\vp}$, 
The $\G_0$ and $\D m$ are chosen to be $\G_0=0.1$~GeV and  $\D m=\{-0.03, 0.01, 0.05\}$~GeV, where $\a_{\rm sh}$'s for these three cases are shown in upper panel of the $1^{\text{st}}$ figure in Fig.~\ref{fig_ansatz}. 
The  $\G_0=0.1$~GeV is not unreasonable large for the particles close to the phase boundary in heavy-ion collisions since even for heavy mesons like $J/\psi$ can have a width around 0.1 GeV in this region~\cite{He:2021zej,Liu:2017qah}. 
For lighter vector mesons, such as $\phi$, it is reasonable to expect a width of this order. 
As shown, depending on the sign and magnitude of a small mass shift (ranging from $-30$ to $50$~MeV), $\alpha_{\rm sh}$ can reach values as large as ``1" and as small as ``$-0.8$". The mass shifts around the phase boundary can be huge~(few hundreds MeV)~\cite{Liu:2017qah,Liu:2016ysz}.  The sign-flipping behavior is mainly due to the competition between the $-2\Delta\ve_{\vp}/\Gamma_{\vp}$ and $\Gamma_{\vp}/2T$ in Eq.~(\ref{eq_ashonly}). 

With these coefficients, the spin alignments  $\rho_{00}$ generated within the framework are shown in the lower panel of the first figure of Fig.~\ref{fig_ansatz}. The magnitude of spin alignment roughly ranges from -2\%  to the 4\%, which is orders of magnitude larger than previous predictions~\cite{Xia:2020tyd} and close to what is observed in experiments~\cite{STAR:2022fan}. Meanwhile, for all three $\D m$ cases, the $p_T$ dependence scales with the $p$ dependence of the coefficients. In particular, for the  $\D m=0.01$~GeV case, the  $p_T$-dependent sign-flipping behaviors observed in experiments are reproduced.  

As a cross check of hydrodynamic calculation, we also estimate the $\rho_{00}$ with $\vp=0$ in the medium rest frame using some simple math following Eq.~(\ref{eq_Tash}):
\begin{align}
	\label{eq_bjo}
	&\d\rho_{00}\propto\frac{1}{9}\alpha_{\text{sh}}\b(\pd_z u^z+\pd_x u^x-2\pd_y u^y)\no\\
	&\d\rho_{00}=\frac{1}{9}\alpha_{\text{sh}}\b\pd_z u^z=\frac{1}{9}\alpha_{\text{sh}}\frac{\b}{\t}, \text{Ideal Bjorken}\,. 
\end{align}
With a positive $\a_{\text{sh}}$ and the fact that usually $\partial_z u^z>\partial_x u^x>\partial_y u^y$, we will get positive spin alignment. Using an ideal Bjorken hydrodynamic solution with $\alpha_{\text{sh}}\sim1$ and $\tau \sim6$~fm,  we will have $\rho_{00}\sim6\%$, which is not very far from the 4\% obtained from a realistic hydrodynamics simulation shown in Fig.~\ref{fig_ansatz}.

\begin{figure} [H]
	\vspace{0.5cm}
\includegraphics[width=0.49\columnwidth]{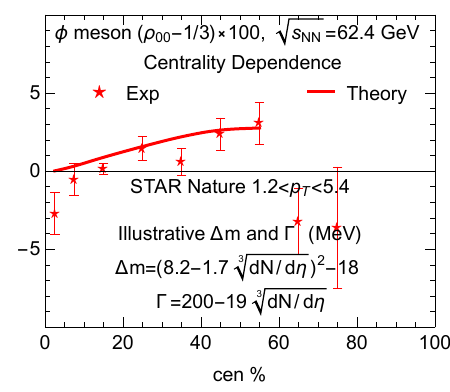}
\includegraphics[width=0.49\columnwidth]{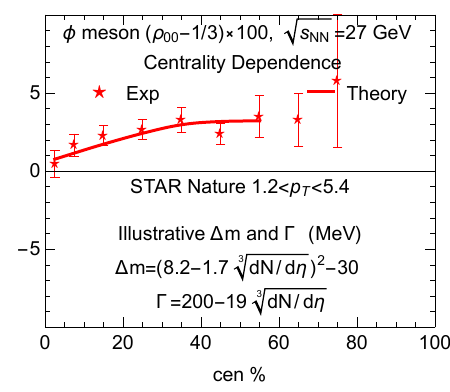}
\includegraphics[width=0.49\columnwidth]{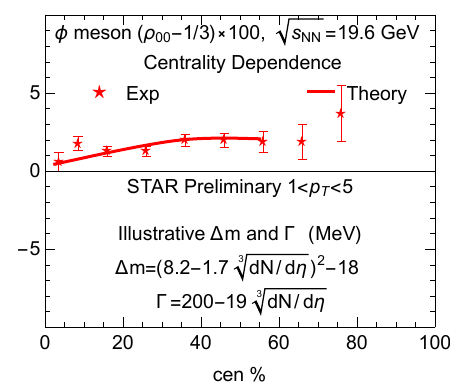}
\includegraphics[width=0.49\columnwidth]{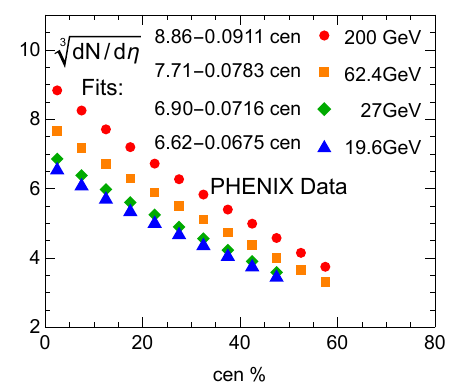}
\vspace{-0.5cm}
\caption{
	Centrality dependence with experimental data~\cite{STAR:2022fan,Wilks:2025uzh} at 62.4~GeV ($1^{\rm st}$ panel), 27~GeV ($2^{\rm nd}$ panel), 19.6~GeV ($3^{\rm rd}$ panel), and $\sqrt[3]{dN/d\eta}$ as a function of centrality from PHENIX data~\cite{PHENIX:2015tbb} ($4^{\rm th}$ panel). 
	Width and mass shift can be fitted as universal functions: $\Delta m=(8.2-1.7\sqrt[3]{dN/d\eta}) +\rm const$ and $\Gamma= 200-19\sqrt[3]{dN/d\eta}$, where ``const" can be fixed by $\rho_{00}$ at 20-60\% for a given beam energy using the same experimental cuts. With the functional form fixed at $200$~GeV, the shape of the centrality dependence at all lower energies is a qualitative ``quasi-prediction" from the proposed ``centrality-freezeout correlation". 
}
\label{fig_cen}
\end{figure}

By correlating beam energies and centralities with spectral properties, such as mass shift or width, we can generate spin alignments with rich behavior. In Fig.~\ref{fig_ansatz}, the $2^{\text{nd}}$ figure demonstrates that a change of less than 40 MeV (ansatzes in figures)  in the in-medium $\phi$ mass with beam energy mimics the observed beam energy dependence of $\rho_{00}$ for $\phi$ and $K^{*}$. The out-plane $\rho_{00}$ is automatically separated from the in-plane one. Although the splitting size depends on the details of the hydrodynamic simulation, the general hierarchy is robust and resembles that observed in experiments. We employ different illustrative ansatzes for the spectral properties of $K^{*}$ and $\phi$, reflecting their different in-medium interactions and freezeout stages: $K^*$ freezes out in the late hadronic phase (positive mass-shifts~\cite{Ilner:2016xqr}), while $\phi$ is assumed to freeze out earlier (QGP or early hadronic phase).The corresponding $p_T$ dependence of $\rho_{00}$ is shown in the $3^{\text{rd}}$ figure, which can accommodate a $p_T$-dependent sign-flipping behavior. 
Correlating spectral properties with $\sqrt[3]{dN/d\eta}$, fitted linearly with centrality~\cite{PHENIX:2015tbb} ($4^{\text{th}}$ figure, more in Fig.~\ref{fig_cen} Supp), can generate a sign-flipping $\rho_{00}$ dependence, mimicking those observed in experiments.

Microscopically, the spectral function's beam energy dependence may relate to the chemical potential, which increases at lower energies. The physics correlating centrality with spectral properties is more subtle. In the next section, we use a microscopic quark-meson model to illustrate this physics further. It is worth noting that  there are still many other potential effects, such as the strong field effect~\cite{Sheng:2022wsy}, initial/local fluctuations ~\cite{Kumar:2023ghs,Fang:2024vds}, etc.~\cite{Kumar:2023ojl,Zhang:2024mhs,Li:2024qae}, that might be entangled. Nevertheless, our calculations support the importance of the thermal-fluctuation-dissipation contribution and  demonstrate that spin alignment is a candidate to serve as a sensitive ``spectrometer", showing that a few-tens-MeV change in mass-shift and width can generate distinguishable behaviors in experimental observables.



\begin{figure} [!t]
	\centering
	\vspace{-0cm}
	\includegraphics[width=0.49\columnwidth]{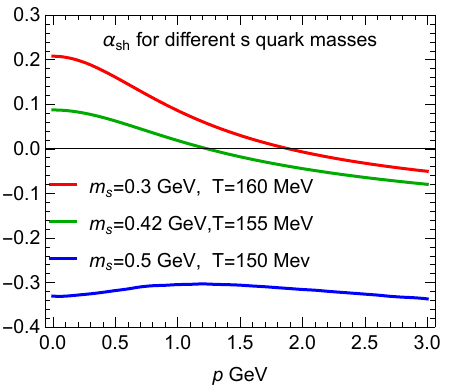}
	\includegraphics[width=0.49\columnwidth]{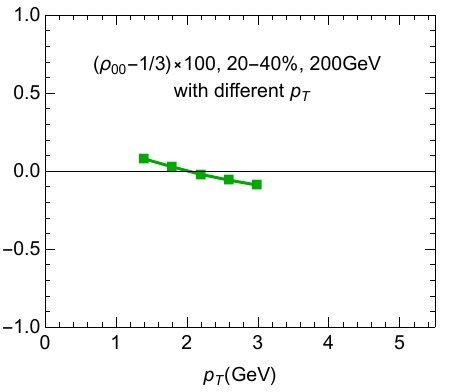}
	\includegraphics[width=0.49\columnwidth]{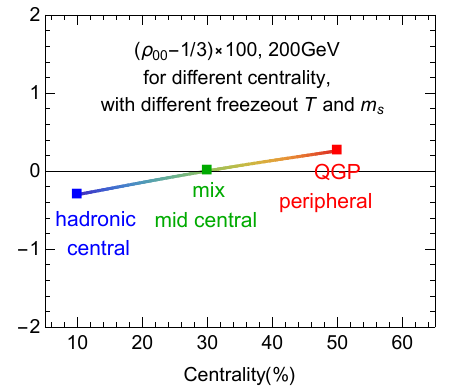}
	\includegraphics[width=0.49\columnwidth]{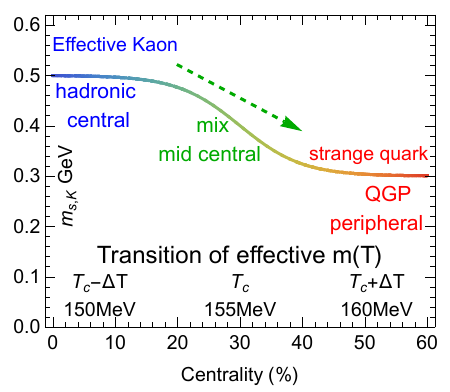}
	\vspace{-0.5cm}
	\caption{
		(1) The $\alpha_{\rm sh}$ coefficients for different cases; (2) $p_T$-dependent $\rho_{00}$ for 'green' coefficients in (1); (3) $p_T$-integrated $\rho_{00}$ as a function of centrality; (4) illustration of $m_s$ change with centrality and freezeout $T$.
	}
	\label{fig_toyrho}
	\vspace{-0.5cm}
\end{figure}

\sect{Quark-meson model coefficients}
\label{sec_QMphen}
To further justify physically motivated coefficients and obtain the model-independent insights, we  have done a microscopic calculation on the coefficients and used it to calculate the spin alignment in this section. We first evaluate the coefficients $\alpha_{\text{sh}}$  numerically using  Eq.~(\ref{eq_ashonly}) for $\phi$ meson with in-medium spectral functions obtained via a 1-loop quark meson model as discussed in Supp (similar ideas in Ref.~\cite{Dong:2023cng}). With coupling  $g_{\phi}=2$, these coefficients are plotted in Fig.~\ref{fig_toyrho} for the cases with the effective strange quark masses $m_s=\{0.3,0.42,0.5\}$~GeV, which are relevant mass scales of strangeness at the QGP phase, mix phase, and hadronic phase. In the hadronic phase, this mass represents a effective in-medium kaon mass (see discussion later). 

As shown in Fig.~\ref{fig_toyrho}, microscopically calculated $\alpha_{\text{sh}}$ generates a $p$ dependence and sign-flipping behaviors similar to those plotted in Fig.~\ref{fig_ansatz}, and they share similar physics but there are some extra physics of the blue curves.  
For $m_s=0.5$~GeV, which is an effective kaon mass,  the $\phi$ meson mass is so close to the $2 m_s$ threshold that $\Gamma_{\vp}$ becomes much smaller, making $\Gamma_{\vp}/(2T) < 2\Delta\varepsilon_{\vp}/\Gamma_{\vp}$ and  $\alpha_{\text{sh}}$ negative for all momenta. This threshold physics will be similar if we switch to a hadronic Lagrangian, as illustrated in our following work~\cite{Sun:2025ror}, where $\alpha_{\text{sh}}$ can be $(-1.5\sim-0.5)$ using a chiral perturbative hadronic Lagrangian or a face value of $\alpha_{\text{sh}}\sim-2$ from Ref~\cite{Haglin:1994ap}.
Meanwhile, we note that the quark meson model used here does not generate large positive $\a_{\rm sh}$. This is because the current medium effects are dominated by radiative effects, and the effective mass shift is  positive, making it difficult to generate large positive values of $\alpha_{\rm sh}$. However, in other situations where collision broadening effects dominate, large widths and negative mass shifts might occur simultaneously~\cite{Liu:2017qah}, leading to larger positive $\alpha_{\rm sh}$. Therefore, in a more realistic situation, we can expect a large positive coefficient similar to those shown in Fig.~\ref{fig_ansatz} in certain scenarios.

The $2^{\rm nd}$ and $3^{\rm rd}$ panels in Fig.~\ref{fig_toyrho} illustrate the $p_T$ dependence and the centrality dependence of the spin alignment from the QM model. For these figures, we actually include all new effects (splitting, zeroth order) discussed in Supp and we confirm that the SITP effect is the dominant one in this setup. As shown,  microscopic and  more complete calculation can recover the sign-flipping $p_T$ and centrality dependence illustrated Fig.~\ref{fig_ansatz}. While the $p_T$ dependence has already been discussed, the``centrality-freezeout correlation" is illustrated in the $4^{\rm th}$ panel  that we utilize hydrodynamic simulations with a different (spin) freezeout temperature $T_{\rm fz}$ correlated with centrality, which is then corrected with $m_s(T_{\rm fz})$ and those $\a_{\rm sh}$ coefficients. 

The plotted ``centrality-freezeout correlation", where $T_{fz}$ increases with centrality (more peripheral), resembles the trend in kinetic freeze-out~\cite{STAR:2017sal}.
The hadronic phase lifetime is largely linear with $\sqrt[3]{dN/d\eta}$~\cite{Knospe:2015nva,Knospe:2021jgt,ALICE:2017ban,ALICE:2020mkb}. Since $\sqrt[3]{dN/d\eta}$ also decreases linearly with centrality (see Fig. ~\ref{fig_cen}), the lifetime drops from 7~fm (0\%) to 2~fm (60\%) at top RHIC.
At large centrality, the shorter hadronic phase, which inhibits re-equilibration in a later, lower-temperature stage, effectively raises the ``freezeout temperature". Meanwhile, the mass-dropping scheme employed for the effective s-quark (or other constituents, such as kaons) is mostly motivated by chiral symmetry restoration physics~\cite{Buballa:2003qv, Li:2016rqo} and emerging low-lying QGP collective modes~\cite{Liu:2017qah} that might be hundreds of MeV below the condensate mass. On the other hand, for those who have concerns about the idea of the strange-masses-melting scenario, we can go back to the Fig.~\ref{fig_ansatz}, where the mass-shifts and width changes required is less than 100~MeV. This is not a strong requirement or unreasonable assumption. Collisional interactions in the QGP phase can generate negative mass shifts as large as a few hundred MeV~\cite{Liu:2017qah}, while radiative ($\phi\rightarrow \bar{K}K  \,\text{or}\, \bar{s}s$ type)  interactions generate positive mass-shifts as shown in our QM model or \cite{Haglin:1994ap}. Therefore, if the interaction type transitions from collisional to radiative type as it crosses the phase boundary, we can expect a sign-flipping in mass shift, leading to sign-flipping in  spin alignment.

\begin{figure} [!tb]
	\centering
	\includegraphics[width=0.99\columnwidth]{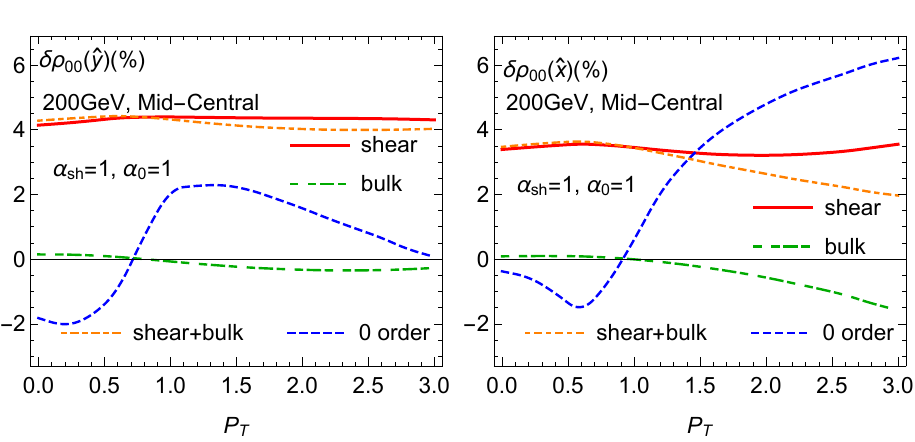}
	\vspace{-0.7cm}
	\caption{
		The effects of the first three contributions (``0 order", ``bulk", ``shear") in Eq.~(\ref{eq_expand}) with $ \hat{y} $ (left) and $\hat{x}$ (right) as quantization axis and $\alpha_0=\alpha_{\text{sh}}=1$(Using unit coefficients to highlight the effects of the flow gradients). The final results should be scaled by the physical value of $\alpha_0$ and $\alpha_{\text{sh}}$.
	}
	\label{fig_example}
	\vspace{0.5cm}
	\includegraphics[width=0.5\columnwidth]{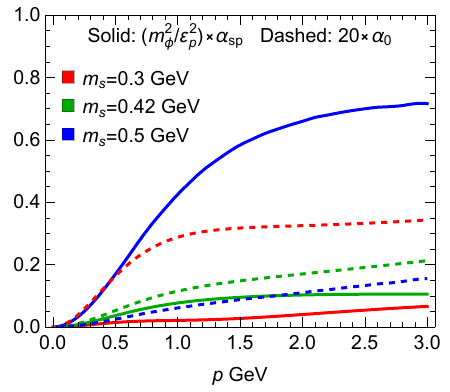}
	\vspace{-0.5cm}
	\caption{Momentum dependence of the ($m_{\phi}^2/\varepsilon_{p}^2)\times \alpha_{\text{sp}}$ and $20\times\alpha_{0}$ with different s quark masses for quark meson model calculations. As has been checked, the effects driven by these coefficients are sub-leading and do not affect the qualitative conclusions in the core part of the paper.
	}
	\label{fig_toyco}
	
	\vspace{0.5cm}
	\includegraphics[width=0.49\columnwidth]{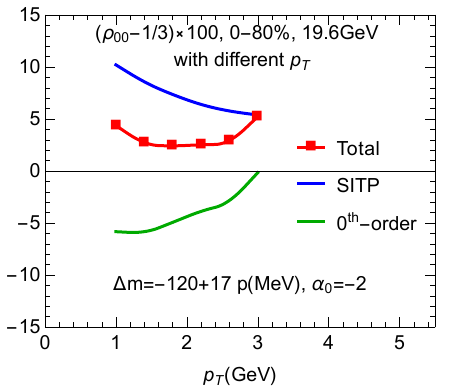}
	\includegraphics[width=0.49\columnwidth]{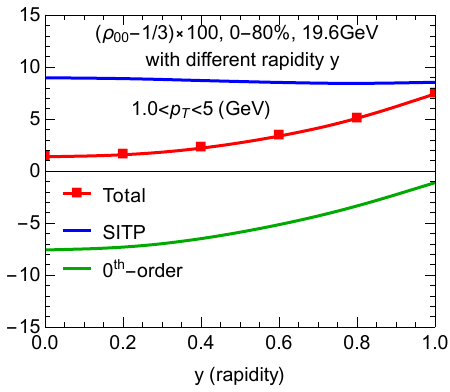}
	\vspace{-0.5cm}
	\caption{An example of the $p_T$ and rapidity $y$ dependence, including the zeroth-order term, with parameters shown in the figures. The $m_{\rm split}$ leads to an $\alpha_0$ on the order of -100 MeV, if we use $\alpha_0 = c_0 m_{\rm split}/T$ for conversion and assume it shares the same $c_0 \sim 2.3$ from the quark-meson model ($m_s = 0.3$).}
	\label{fig_ypt}

\end{figure}

Besides published data, preliminary experimental data~\cite{Wilks:2025uzh} suggest a strong rapidity dependence at lower energies. While the SITP effect is largely boost-invariant at mid-rapidity, this dependence can be accounted for by including the zeroth-order term with coefficient $\alpha_0$ (see Eq.~(\ref{eq_T0}) Supp). Tuning both $\alpha_0$ and $\alpha_{\rm sh}$ can accommodate this new feature without qualitatively altering the features shown in Fig.~\ref{fig_ansatz}. A large rapidity dependence would imply a $\phi$ meson with a longitudinal/transverse mass splitting at the order of 100 MeV (see  large splitting in other particles~\cite{Yeo:2024iqz}), a possible case for a strongly-coupled medium (see Fig.~\ref{fig_ypt}  Supp).

\sect{Summary}\label{sec3}
We have developed a Zubarev response approach based on thermal field theory and utilized it to study the tensor polarization and spin alignment of vector mesons, leading to two key theoretical findings:

\noindent(1) We derive for the first time a fluctuation-dissipation relation for tensor polarization and spin alignment, revealing spin alignment is naturally related to damping properties and, consequently, to the spectral functions.

\noindent(2) We discover a novel shear-induced tensor polarization (SITP) effect at first order of hydrodynamic gradients, which can generate significant spin alignment.

Phenomenologically, by tuning the parameter $\alpha_{\rm sh}$, the SITP effect can generate spin alignment that depends on beam energy, centrality, $p_T$, quantization axes, and particle type ($\phi$, $K^*$), resembling those observed in experiments, including various sign-flipping behaviors. Further including zeroth-order effect with an extra parameter, $\alpha_0$, can produce a large rapidity dependence. 

In the context of the symmetry-based gradient expansion (see Eq.~(\ref{eq_expand})), this study demonstrates that terms proportional to $\pd^{(\m} \x^{\n)}$ and $u^\m u^\n$ already provide strong explanatory power for current experimental observations.  This offers a promising path toward understanding spin alignment physics within the general framework of the relativistic hydrodynamic model, often referred to as the ``standard model" in heavy-ion physics.

Meanwhile, we also demonstrate that a few-tens-MeV change in mass-shifts and width can significantly change the transport coefficients associated with the above gradients, generating distinguishable behaviors in experimental observables and making spin alignment serve as a promising ``spectrometer" candidate for vector mesons.

To fully realize this ``spectrometer" blueprint, we need to investigate additional contributions beyond the leading orders to refine the relation between the spectral function and spin alignment. This effort also requires more sophisticated studies of in-medium spectral properties, alongside further improved phenomenology and simulations. Investigating other non-equilibrium effects and novel contributions proposed in Refs.~\cite{Sheng:2022wsy,Muller2021} is also necessary to understand their quantitative impact. Finally, as a separate point, the newly discovered SITP effect is robust against variations in interaction details and is universal in both relativistic and non-relativistic cases, making it interesting to investigate experimentally in low-energy physics, such as plasma or cold atom physics~\cite{ColdAtom}.

\acknowledgments
\textbf{Acknowledgments}
We gratefully acknowledge the valuable contribution from Yi Yin, who definitely deserves a key position in the author list, but waive the authorship generously. 
We extend special thanks to Xiangyu Wu for sharing the most recent CLVisc code and providing helpful guidance for our simulations at various beam energies in later versions. We are also especially grateful to Longgang Pang for providing hydrodynamic profiles for our early numerical studies.
We also thank Youyu Li, Xinxiang Liu, and Tan Luo for engaging in valuable discussions, digitizing experimental plots, and providing guidance on simulations.
We thank Hengtong Ding, Xiaojian Du,  Aiqiang Guo, Koichi Hattori, Min He, Che-Ming Ko, Yutie Liang, Shu Lin, Rapp Ralf, Kaijia Sun, Subhash Singha, Shuzhe Shi, Yifeng Sun, Aihong Tang, Biaogang Wu, Dilun Yang, Wenbin Zhao
for valuable discussions.
SL is supported by NSFC No. 12205090. Fundamental Research Funds for the central Universities.
FL is supported by NSFC No. 12105129.

\bibliography{ref}
\pagebreak

\section{supplemental material (Supp)}
\subsection{Theoretical derivation for tensor polarization}
In this section, we will provide more details on the derivation that leads to the theoretical results presented in the previous sections. We  begin by introducing the \textit{Wigner functions} of a massive vector meson with mass $m$ and field operator $V^\mu$:
\begin{align}
	\label{eq_Wdefine}
	W^{\mu\nu}(x,\vp)&\equiv\varepsilon_{\vp}\int \frac{dp^0}{2\pi} \int {d^4 y}e^{i p\cdot y}\langle V^{\mu}(x_-)V^{\nu}(x_+)\rangle\no\\
	& = W^{\mu\nu}_+(x,\vp)+ W^{\mu\nu}_-(x,\vp)
\end{align}
where $x_{\pm}=x\pm y/2$, $\ve_{\vp}=\sqrt{\vp^{2}+m^{2}}$, $\langle\ldots\rangle$ denotes the thermal ensemble average, and $W_\pm$ denote the integration over the positive and negative $p^0$ respectively, corresponding to the differential spin density matrix $\varrho(x,\pm\vp)$~\cite{Becattini:2013fla}, where
the $\varrho(x,\vp)$ is embedded in Wigner function as
\begin{align}
	2 W^{\mu\nu}_+(x,\vp) &= \sum_{s,s^\prime}\epsilon^{\mu\ast}_s(\vp) \epsilon^\nu_{s^\prime}(\vp) \varrho_{s^\prime s} (x,\vp) +\delta W^{\mu\nu}\no\\
	&\equiv \mathcal{W}^{\mu\nu}(x,\vp)+\delta W^{\mu\nu}(x,\vp)\,.
\end{align}
The factor ``2" is for matching the conventional normalization of the density matrix.
The $\epsilon_s(\mathbf p)$ represents the polarization vector of the vector meson moving with momentum $\vp$ and occupying the spin state $s$, and satisfies $\tilde{p}\cdot \epsilon_s(\vp) = 0$ with $\tilde{p} = (\varepsilon_{\vp}, \vp)$ being the on-shell 4-momentum. So, the projected Wigner function
\begin{align}
	\mathcal{W}^{\mu\nu}(x,\vp)\equiv \sum_{s,s^\prime}\epsilon^{\mu\ast}_s(\vp) \epsilon^\nu_{s^\prime}(\vp) \varrho_{s^\prime s} (x,\vp)
\end{align}
is perpendicular to the on-shell 4-momentum $\tilde p$ as well, and can therefore be expressed as
\begin{align}
	\mathcal{W}^{\mu\nu} (x,\vp) \equiv 2\tilde\Delta^\mu_\alpha \tilde\Delta^\nu_\beta W^{\a\b}_+ (x,\vp)
\end{align}
where $\tilde\Delta$ is the shorthand of $\Delta(\tilde p)$ with $\Delta^{\mu\nu}(p)\equiv -\eta^{\mu\nu} + p^\mu p^\nu /p^2$ being the projector with respect to a 4-momentum $p$ (note $\Delta^2=-\Delta$). The $\d W$ can always be  chosen vanishing after projections. Conversely, the differential spin-density matrix can be evaluated via the projected Wigner function as (There are subtleties on this relation, but they will not affect symmetric parts of $\mathcal{W}^{\mu\nu}$ \cite{Hattori2020})
\begin{align}
	\label{eq_rhofW}
	\varrho_{ss^\prime} (x,\vp)= \epsilon_{s'}^\mu(\vp) \epsilon^{\nu\ast}_{s}(\vp) \mathcal{W}_{\mu\nu}(x,\vp).
\end{align}

The projected Wigner function can be further decomposed, according to the representation of the rotational symmetry, into three parts as
\begin{align}
	\label{eq_Wsvt}
	&\sW^{\mu\nu}
	=\frac{1}{3}\tilde{\Delta}^{\mu\nu} \sS+\sW^{[\mu\nu]}+ \mathcal{T}^{\mu\nu}
\end{align}
where $ \sS\equiv\sW^{\mu\nu}\tilde{\Delta}_{\mu\nu} $ is the 3D trace of $\sW^{\mu\nu}$ and related to spin-summed phase space density, $ W^{[\mu\nu]}\equiv (W^{\mu\nu}- W^{\nu\mu})/2$
corresponds to the vector polarization of the vector meson, and 
$ \mathcal{T}^{\mu\nu} $ defined as
\begin{align}
	\label{eq_Tdefine}
	\mathcal{T}^{\mu\nu}\equiv \sW^{\langle\mu\nu\rangle}\equiv\sW^{(\mu\nu)}-\frac{1}{3}\tilde\Delta^{\mu\nu} \sS
	=2\tilde\Delta^{\langle\mu}_{\;\l}\tilde\Delta^{\nu\rangle}_{\;\g}W^{(\l\g)}_+
\end{align}
corresponds to the tensor polarization of the vector field, which is of the major interest in this work. Here, the round bracket \lq\lq$\scriptstyle(\cdots)$\rq\rq  stands for symmetrizing the included space-time indices,  i.e., $W^{(\mu\nu)}=(W^{\mu\nu}+ W^{\nu\mu})/2$, while the angle bracket \lq\lq$ \scriptstyle\langle...\rangle $\rq\rq stands for further making the tensor traceless about the included indices. 

$\mathcal T^{\mu\nu}$ can be further expanded in terms of the hydrodynamic gradients as
\begin{align}
	\mathcal T^{\mu\nu} \approx \mathcal T^{\mu\nu}_{(0)} + \mathcal T^{\mu\nu}_{(1)}  = 2 \tilde\Delta^{\langle\mu}_{\;\l}\tilde\Delta^{\nu\rangle}_{\;\g} \left(W_{+(0)}^{(\l\g)}+W_{+(1)}^{(\l\g)}\right)
	\label{eq:TotalT}
\end{align}
where the subscripts $(0)$ and $(1)$ stand for the zeroth and first order of the hydrodynamic gradients (or $\partial$). After listing all the symmetry-allowed  tensor structures non-vanishing under the projection $\tilde\Delta^{\langle\mu}_{\;\l}\tilde\Delta^{\nu\rangle}_{\;\g}$ up to the order of $\partial$, we further express $\sT^{\mu\nu}$ schematically as
\begin{align}
	\label{eq_expand}
	\mathcal T^{\mu\nu}
	=&\tilde\Delta^{\langle\mu}_{\;\l}\tilde\Delta^{\nu\rangle}_{\;\g}\left[\k^{u}_0 u^\l u^\g+\k^{u}_1 u^\l u^\g + \k_{\rm sh}\sigma^{\l\g} +\k_T u^{(\l}\partial_\perp^{\g)}\beta\right.\nonumber\\
	&\left.
	+\k_{\text{su}} u^{(\l}\s^{\g)\a}\tilde{p}_{\a}+\k_{\text{ou}} u^{(\l}\O^{\g)\a}\tilde{p}_{\a} 
	+\cdots\right], 
\end{align}
where $u^\mu$ is the flow velocity and $\beta=1/T$ is the inverse temperature.  $\pd^{\mu}_{\perp}\equiv \bar{\Delta}^{\mu\nu} \partial_\nu$ is the transverse derivative with the flow projector $\bar{\Delta}^{\mu\nu}\equiv \eta^{\mu\nu} -u^\mu u^\nu$. The shear stress tensor is defined as $\sigma^{\mu\nu}\equiv \pd^{(\mu}_{\perp}u^{\nu)}_{\white{\perp}}-(1/3)\bar{\Delta}^{\mu\nu}\theta$ with bulk stress $\theta\equiv\partial \cdot u$.  $\O^{\mu\nu}\equiv \pd^{[\mu}_{\perp}u^{\nu]}_{\white{\perp}}$ is vorticity. The SITP contribution $\k_{\rm sh}\sigma^{\l\g}$ appears naturally in this symmetry analysis with a T-odd coefficient $\k_{\rm sh}$ that should be originated from the dissipative processes. 
In the following, the $\k$-coefficients will be evaluated near thermal equilibrium under the framework of thermal field theory and linear response theory, with inclusion of dissipative physics.

The \textit{zeroth-order term} $\mathcal T^{\mu\nu}_{(0)}$ in  Eq.(\ref{eq:TotalT}) can be evaluated under exact/global thermal equilibrium and is related to the in-medium spectral function \lq\lq$A$\rq\rq via
\begin{eqnarray}
	\mathcal T^{\mu\nu}_{(0)} &=& 2\tilde{\Delta}^{\langle\mu}_{\a}\tilde{\Delta}^{ \nu\rangle}_{\b} \int_0^\infty d p^0 \int {d^4 y}e^{i p\cdot y}\langle V^{\alpha}(x_-)V^{\beta}(x_+)\rangle\nonumber\\
	&=& 2 \tilde{\Delta}^{\langle\mu}_{\a}\tilde{\Delta}^{ \nu\rangle}_{\b} \int_0^\infty d p^0  n(p^0) A^{\alpha\beta}(p),
\end{eqnarray}
where $n(\o)=1/(e^{\b\o}-1)$ is the Bose-Einstein distribution. In the thermal medium, the longitudinal and transverse modes of the vector meson are different, so that the spectral function can be decomposed as~\cite{Gale:1990pn,Rapp:1997fs}
\begin{align}
	\label{eq_spec}
	&A^{\mu\nu}=\sum_{a=L,T}\Delta^{\mu\nu}_{a}A_{a}, \,A_{a}=\frac{1}{\pi}\text{Im }\frac{-1}{p^2-m^2-\Pi_a}.
\end{align}
The longitudinal and transverse projector $ \Delta_{T,L} $ are  $ \Delta^{\mu\nu}_{L}=v^\mu v^\nu/(-v^2)$, 
$ \Delta^{\mu\nu}_{T}=\Delta^{\mu\nu}-\Delta^{\mu\nu}_{L}$, where
$v^\mu=\Delta^{\mu\nu}u_\nu$ is the projected flow velocity with respect to $p$. The on-shell version of these projectors are denoted by $\tilde{\Delta}^{\mu\nu}_{L}=\tilde{v}^\mu \tilde{v}^\nu/(-\tilde{v}^2)$ with $\tilde{v}^\mu =\tilde{\Delta}^{\mu\nu}u_\nu$ and $\tilde{\Delta}^{\mu\nu}_{T}=\Delta^{\mu\nu}_{T}$. Given that $\tilde\Delta^{\mu}_{\l}\tilde\Delta^{\nu}_{\g}\Delta^{\l\g}_{L}(p)=\tilde{\Delta}^{\mu\nu}_{L}(1-\D \o^2 \tilde{v}^2 /p^2)$ with $\D\o=p^0-\ve_{\vp}$,
we therefore obtain that 
\begin{align}
	\label{eq_T0}
	&\sT^{\mu\nu}_{(0)}=\a_\text{0}n(\ve_{\vp})\tilde{\D}^{\langle\mu\nu\rangle}_L =\frac{\a_0}{-\tilde{v}^2} n(\ve_{\vp}) \tilde{\Delta}^{\langle\mu}_\l\tilde{\Delta}^{\nu\rangle}_\g u^{\l}u^{\g},\no\\
	&\a_\text{0} = 2\ve_{\vp}\int_0^\infty d\o\, \frac{n(\o)}{n(\ve_{\vp})} \Big{[}(A_L-A_T)-\frac{\D\o^2\tilde{v}^2}{p^2}A_L\Big{]}.
\end{align}
$A_L$ usually has a factor proportional to $p^2$ cancelling with $ 1/p^2 $ to make the integration convergent. 
In principle, the $\a_0$ in Eq.~(\ref{eq_T0}) should be evaluated numerically using a realistic in-medium spectral function.  However, if we assume a non-analytic real energy shift at zero width limit, $\a_0\approx(\o^T_{\vp}-\o^L_{\vp})/T$ where  $\o_{\vp}^{L/T}$ satisfying  $ (\o^{L/T}_{\vp})^2-\ve_{\vp}^2-\text{Re}\Pi_{L/T}(\o_{\vp}^{L/T},\vp)=0$ represents the shifted dispersion relation of the longitudinal and the transverse modes, respectively. 

In the medium rest frame, nonzero momentum $\vp$ breaks the rotational symmetry, which results in the difference between $A_L$ and $A_T$ and therefore leads to the tensor polarization even in the absence of hydrodynamic gradient
as expressed by the first term in $\a_0 $ in Eq.~(\ref{eq_T0}). Such splitting induced polarizations have also been discussed for virtual photons~\cite{Baym:2017qxy,Speranza:2018osi}, but this is the first time it has been discussed in the context vector meson in a locally equilibrated hydrodynamic medium. 
In addition, the projection of an off-shell in-medium particle mismatches with the on-shell final states particle, which leads to the second term in $\a_0 $ in Eq.~(\ref{eq_T0}). 

We then turn to the \textit{first-order term} $\mathcal T^{(1)}$, which is proportional to the hydrodynamic gradients and accounts for the  off-equilibrium contribution induced by the system inhomogeneity.

Using the Zubarev's formalism~\cite{Hosoya:1983id,zubarev1974nonequilibrium}, we obtain a Kubo formula at vanishing chemical potential that
\begin{align}
	\label{eq_kubo}
	W_{+(1)}^{\mu\nu}&=\ve_{\vp}\lim_{\omega,q\rightarrow0}\frac{\partial}{\partial\omega}[-\text{Im}G^{\mu\nu \l\g}_{R+}(\omega,\vq,\vp) ]\xi_{\l\g}
\end{align}
where the $G_{R+}$ is embedded in retarded Green function
\begin{align}
	\label{eq_def_Gr}
	G^{\mu\nu\l\g}_R(t-t',\vx,\vz,\vy)\equiv&\int \frac{d\omega}{2\pi} \frac{d^3\vq}{(2\pi)^3} \frac{d^3\vp}{(2\pi)^3}e^{-i\omega\cdot(t-t')}
	\no\\
	\times& e^{i\vq\cdot(\vx-\vz)}e^{i\vp\cdot\vy}G^{\mu\nu \l\g}_{R}(\omega,\vq,\vp)
	\no\\
	=(-i)\Theta(t-t')\langle [V^\mu(t,&\vx_{-}) V^{\nu}(t,\vx_{+}),T^{\l\g}(t',\vz)]\rangle,
\end{align}
and $\xi_{\l\g} \equiv \beta^{-1} \partial_{(\l} (\beta u)_{\g)}$.
Combined with the ideal hydrodynamic equations, $\xi_{\l\g}$ can be written as a linear combination of $\sigma_{\l\g}$ and $\theta$ at the leading gradient order as
\begin{equation}
	\label{eq:HydroDerivative}
	\xi_{\l\g}\approx \sigma_{\l\g} + \left[\frac {1}{3} \bar{\D}_{\l\g}+c_s^2 u_\l u_\g\right]\theta,
\end{equation}
where
$c_s^2$ is the square of the sound speed, whose value at chemical freeze out, taken from lattice QCD data~\cite{Borsanyi:2013bia}, is around 0.16. The shear part of the Kubo formula can also been obtained using metric variation in Ref~\cite{Luttinger1964,Moore:2010bu}. 

In Eq.~(\ref{eq_def_Gr}), $T^{\mu\nu}\equiv -F^{\mu}_{\;\;\a}F^{\nu\a}+m^{2}V^{\mu}V^{\nu}- \eta^{\mu\nu}\le(-F^{2}/4+m^{2}V^{2}/2\ri)$  stands for the Belinfante energy-momentum tensor of the sourceless Proca field, where $F^{\mu\nu}\equiv\pd^{\mu}V^{\nu}-\pd^{\nu}V^{\mu}$.
After keeping only the leading order of the skeleton expansion\cite{Srednicki:2007qs,Dyson:1949ha,PhysRev.118.1417}, we express the $ G^{\mu\nu \l\g}_{R}(\omega,\vq,\vp) $ in the medium rest frame as:
\begin{align}
	\label{eq_G_byDp} 
	G^{\mu\nu \l\g}_{R+}(\omega,\vq,\vp)=&-\int^{\infty}_0 dk_0\int^{\infty}_0 dk_0'\frac{n(k_0')-n(k_0)}{\omega+k_0'-k_0+i0^+}
	\no\\
	\times\sum_{a,b=L,T}&A_{a}(k)A_{b}(k')I^{\mu\nu\l\g}_{ab}(k,k')\,.
\end{align}
The integral limits excluding the negative energy region allow us to conveniently select out the modes that are related to the physical modes in the Wigner functions. The neglected contribution is of high orders of $\delta_{\text{qp}}$ defined later in the quasi-particle approximation schemes. With the projectors, the $ I^{\mu\nu\l\g}_{ab}(k,k')$ in Eq.~(\ref{eq_G_byDp}) can be explicitly written as
\begin{align}
	\label{eq_I_cal} 
	&I^{\mu\nu\l\g}_{ab}(k,k')=	[k^{\l} k^{\prime\g}+ k^{\g} k^{\prime\l}]\D^{\nu\a}_{a}(k)\D^{\mu}_{b,\a}(k')
	\no
	\\
	&-[k_{\a} k^{\prime\g}\D^{\nu\l}_{a}(k)\D^{\mu\a}_{b}(k')+k^{\g} k^{\prime}_\a\D^{\nu\a}_{a}(k)\D^{\mu\l}_b(k')]
	\no\\
	&-[ k^{\l}k^{\prime}_{\a}\D^{\nu \a}_{a}(k)\D^{\mu \g}_{b}(k')+k_{\a}k^{\prime\l}\D^{\nu \g}_a(k)\D^{\mu \a}_b(k')]
	\nonumber\\
	&+(k_{\a} k^{\prime \a}-m^2)[\D^{\nu \l}_a(k)\D^{\mu \g}_{b}(k')+\D^{\nu \g}_{a}(k)\D^{\mu \l}_{b}(k')]
	\no\\
	&-\eta^{\gamma \l}[(k^{\zeta}k^{\prime}_{\zeta}-m^2)\eta_{\a\b}-k_{\b}k^{\prime}_{\a}]\D^{\nu\a}_{a}(k)\D^{\mu\b}_{b}(k')]
\end{align}
where the $ k=(k_0,\vp+\vq/2) $, $ k'=(k_0',\vp-\vq/2) $. 

To proceed further, we carry out the integration in the quasi-particle limit, i.e., $\Pi_{T,L}/\ve_{\vp}^2\sim\d_{\text{qp}} \ll 1$, so that all the integral over $ k_0 $ and $ k_0' $ can be done in the vicinity of the pole of the propagator. Also, we assume the splitting is small, i.e.,  $(\Pi_{L}-\Pi_{T})/(\Pi_{L}+\Pi_{T})\sim\d_{\text{sp}} \ll 1$. Up to the (overall) zeroth order of $\d_{\G}$ and $\d_{\text{sp}}$, the off-shell contribution in the projectors in Eq.(\ref{eq_G_byDp}) can be neglected, and $	\sT^{\mu\nu}_{(1)}$ can be simplified as
\begin{align}
	\label{eq_T1full}
	\sT^{\mu\nu}_{(1)}
	=\beta n(\ve_{\vp}) \tilde\Delta^{\langle\mu}_\l\tilde\Delta^{\nu\rangle}_\g \big{[}  \alpha_{{\rm sh}} \xi^{\g\l}  +\a_\text{sp}\xi_{p}\frac{u^\l u^\g}{-\tilde v^2}\big{]}
\end{align}
where $\xi_{p}=(\tilde p^\rho \tilde p^\sigma)\xi_{\rho\sigma}/\ve_{\vp}^2 $ and
\begin{align}
	\label{eq_ash}
	\alpha_{\rm sh}=& \frac{4\ve_{\vp} \pi}{ \beta n(\ve_{\vp})}\int^\infty_0d\o\frac{\partial n(\o)}{\partial \o}  (\o^2-\ve_{\vp}^2) A^2_{T/L}(\o,\vp)\\
	\a_\text{sp}=& \frac{4\ve_{\vp} \pi}{\beta n(\ve_{\vp})}\int^\infty_0d\o\frac{\partial n(\o)}{\partial \o} \ve^2_{\vp} (A_T^2(\o,\vp)-A_L^2(\o,\vp))\,.\no
\end{align}
Furthermore, in quasi-particle limit, the spectral function nearby the positive frequency pole can be approximately expressed as 
\begin{equation}
	\label{eq_quasiA}
	A_a(\omega,\vp) \approx \frac{1}{2\ve_{\vp}}\frac{1}{\pi} \mathrm{Im} \frac{-1}{\omega - \omega^a_{\vp}+i \Gamma^a_{\vp}/2}
\end{equation}
where $\Gamma_{\vp}^{a} =\text{Im}\Pi_{a}(\o^{a}_{\vp},\vp)/\ve_{\vp} $ is the width.  With an expansion of $\pd n(\o)/\pd\o$ to the first order of $\D\o/T$ as $\propto n(\ve_{\vp})(1-\D\o/T)$, 
$\alpha_{\rm sh}$ and $\alpha_{\rm sp}$ can be further simplified as with  identities $ \int dx\,x^2 (c/((x)^2+c^2))^2=\pi c/2$, $ \int dx (c/(x^2+c^2))^2=\pi/(2c)$:
\begin{align}
	\label{eq_ash2}
	&\alpha_{\rm sh} \approx
	-\frac{2\Delta\ve_{\vp}}{\Gamma_{\vp}}+2\frac{\Delta\ve_{\vp}}{\Gamma_{\vp}}\frac{\Delta\ve_{\vp}}{T}+\frac{\Gamma_{\vp}}{2T} \sim \mathcal O(1)\\
	&\a_\text{sp}
	\approx-\frac{\ve_{\vp}}{\G_{\vp}}\left(\frac{\G^{\D}_{\vp}}{\G_{\vp}}-\frac{\D\ve_{\vp}}{T}\frac{\G^{\D}_{\vp}}{\G_{\vp}}+\frac{\G_{\vp}}{T}\frac{\o^{\D}_{\vp}}{\G_{\vp}}\right) \sim \mathcal O(\delta_{\rm qp}^{-1} \delta_{\rm sp}).\no
\end{align}
with $\G^{\D}_{\vp} \equiv\G^{L}_{\vp}-\G^{T}_{\vp}$ and $\o^{\D}_{\vp} \equiv\o^{L}_{\vp}-\o^{T}_{\vp}$ being the difference between the width and dispersion relations of the $L$ and $T$ modes. 
$\Delta \ve_{\vp}$ and $\G_{\vp}$ are defined as $\Delta \ve_{\vp}=\o_{\vp}^{L/T}-\ve_{\vp}$ and $\G_{\vp}=\G^{L/T}_{\vp}$, where the differences caused by choosing $L/T$ are  $\mathcal O(\d_{\text{sp}})$.
Error of the expansion is around 20\%, even for relatively large $\D\o/T$. For $\G_{\vp}/T(\ve_{\vp}/T)\sim0.5$. Also, the Bose enhancement factors in Eq.~(\ref{eq_ash2}) are neglected since $n(\ve_p)$ is small for vector boson with large mass. 
Finally, to make the obtained $\alpha_{\rm sh}$ and $\alpha_{\rm sp}$ covariant, we need to replace $\vp$ with $ p_{\perp}^\mu\equiv \bar{\Delta}^{\mu\nu} p_\nu $ and $ \ve_{\vp} $ with $\ve_{0}\equiv\tilde p\cdot u $.

After comparing Eq.~(\ref{eq_T0}) and Eq.~(\ref{eq_T1full}) with Eq.(\ref{eq_expand}), we obtain the $\kappa$-parameters at current order as
\begin{align}
	\label{eq_kappa}
	&\k^u_0 =\frac{\a_0}{-\tilde{v}^2}n_0,~\kappa^u_1 =\left[\alpha_{\rm sh}\left( c_s^2-\frac{1}{3}\right)\theta+\frac{\alpha_{\rm sp}\xi_{p}}{-\tilde v^2}\right] \beta n_0 \nonumber\\
	&\kappa_{\rm sh} = \alpha_{\rm sh} \beta n_0,\quad \kappa_T=0,\quad \kappa_{\text{su}}=0,\quad\kappa_{\text{ou}}=0
\end{align}
with $n_0=n(\ve_0)$. The $\a_0$ is at least of first order in $\d_{\text{sp}/\text{qp}}$ but of zeroth order in $\pd$. Thus, $\sT$ in this multi-parameter expansion is at the first order of $\pd$, $\d_{\text{sp}}$ or $\d_{\text{qp}}$.
For higher orders, there are non-vanishing contributions from $\k_{\text{su}}\propto \d_{\text{sp}/\text{qp}}$ and $\k_{T}\propto$ ``high order gradients", while $\k_{\text{ou}}$ could be nonzero with asymmetric $T^{\mu\nu}$.

Some relevant and interesting physics other than tensor polarization can be studied under the similar formalism.
For example, the vector polarization for vector boson can be proven equal to 4/3 of the one for spin-1/2 particle as expected. However, due to the limited space, we would leave such discussions together with more details of the present work to our long paper~\cite{Li:2025pef}.

\subsection{More for quark-meson models}
\label{sec_QMtheo}
In this section, we provide some more information on the calculation of the quark meson model discussed in Sec.~\ref{sec_QMphen}, where details will be included in the future works. We begin with Lagrangian of quark meson model that including only one strange quark as
\begin{align}
	L=\int d^4x\,\bar{\psi}(i\slashed{D}-m)\psi-\frac{1}{4}F^{\mu\nu}F_{\mu\nu}+\frac{1}{2}m^2V^\mu V_\mu
\end{align}
where the $D_{\mu}=\pd_{\mu}-i g_{\phi }V_{\mu}$ and $F_{\mu\nu} =\partial_\mu V_\nu-\partial_\nu V_\mu$ with $V_{\mu}$ as $\phi$ meson field. For this Lagrangian, with $k=q+p$, the one-loop self-energy diagram in Matsubara formalism can be expressed as
\begin{align}
	&\Pi^{\mu\nu}(p)=-g_\phi^2 \beta^{-1}\sum_{n}\int\frac{d^3 \vq}{(2\pi)^3}\text{Tr}\{\gamma^{\mu}\frac{\slashed{q}+m_s}{q^2-m_s^2}\gamma^{\nu}\frac{\slashed{k}+m_s}{k^2-m_s^2}\} .
\end{align}
Then, we proceed to sum Matsubara frequency using contour integral,  take the trace and use the symmetry $(q\rightarrow -k, k\rightarrow-q)$ to combine and simplify terms~(details see future works). In the end,  we get the medium part (subtracting the 1 from the vacuums) of the self-energy  be expressed as
\begin{align}
	\Pi^{\mu\nu}_{me}(p)=&-4g_\phi^2\int d q_0 \frac{d^3\vq}{(2\pi)^3}(2\,\text{sign}(q_0) f(|q_0|)) \rho_s(q_0,\bm{q})\no\\&\times\frac{q^\mu k^{\nu}+k^\mu q^\nu-g^{\mu\nu}( q\cdot k-m_s^2)}{k^2-m^2_s}
\end{align}
with $\rho_s(q_0,\vq)=(\delta(q_0-\ve_{\vq})-\delta(q_0+\ve_{\vq}))/(2\ve_{\vq})$,  $\ve_{\vq}=\sqrt{m_s^2+\vq^2}$ and $p$ understood with a $p_0\pm i \epsilon$ 
for analytical continuation. The angular integral can be performed analytically, left one dimensional $q$ integral. It is straight forward to verify $p_\mu \Pi ^{\mu\nu}_{me}=(k_\mu -q_\mu)\Pi ^{\mu\nu}_{me}=0$ from the above formula. 

For the vacuum part $\Pi ^{\mu\nu}_{vac}=(p^2g^{\mu\nu}-p^\mu p^\nu)\pi(p)$,  we just take the standard dimensional regularization part
\begin{align}
	\pi(p)&=-8\frac{g^2_\phi}{(4\pi)^2}\int_{0}^{1}dx\,x(1-x)\ln(\frac{|m^2_s-(1-x)x m_\phi^2|}{m^2_s-(1-x)x p^2})
\end{align}
with wave function and mass counter terms $\delta_A$ and $\delta_m$ to make the numerator $
(p^2(1-\pi(p^2)-\delta_A)-m^2-\delta_m)^{-1}$ having the correct mass and residues. 
\begin{align}
	\delta_A&=8\frac{g_\phi^2}{(4\pi)^2}\text{PV}\int_0^{1}dx\frac{x(1 -  x) (1 - 
		x) x m_\phi^2}{m_s^2 - (1 - x) x m^2_\phi}\no\\
	\delta_m&=-m_\phi^2\delta_A
\end{align}
with PV denoted the principle value integral. For the Matsubara formalism, the above formula set is all we need to reproduce the results in \cite{Dong:2023cng}, where the numerical results are explicitly checked and they are agree with same setup.

\subsection{Backup figures for phenomenology}
This section presents three supplementary figures for reference. These figures either appeared in a previous version of this work or provide additional details.

\end{document}